\documentclass[12pt]{article}

\usepackage{sbc-template}
\usepackage{graphicx,url}
\usepackage[utf8]{inputenc}
\usepackage[T1]{fontenc}
\usepackage[english]{babel}
\usepackage{amsmath,amssymb}
\usepackage{tikz}
\usepackage{placeins}
\usetikzlibrary{arrows.meta,positioning}

\graphicspath{{fig/}}
\title{A Dynamic Vertical Scaling Strategy for Distributed Stream Processing Applications in Edge Computing}

\author{Guilherme Hiago Costa dos Santos\inst{1} Carlos Henrique Kayser\inst{2} Tiago Coelho Ferreto\inst{3}}

\address{Polytechnic School -- Pontifical Catholic University of Rio Grande do Sul
  (PUCRS)\\
  Post Office Box 1429 -- 90.619-900 -- Porto Alegre -- RS -- Brazil
  \email{\{guilherme.hiago,carlos.kayser\}@edu.pucrs.br, tiago.ferreto@pucrs.br}
}

\begin{document}

\maketitle

\begin{abstract}
Distributed Stream Processing applications at the edge must reconcile low latency and high throughput with limited and heterogeneous resources. This paper presents a dynamic vertical scaling strategy based on Proximal Policy Optimization, formulated as a Partially Observable Markov Decision Process. The policy jointly adjusts task allocations and prioritizes compliance with a p95 end-to-end latency Service Level Objective. In EdgeStreamPy simulation experiments with two application profiles, two workloads, and ten paired placements per combination, the selected policies preserved throughput, obtained mean violation rates from 0.03\% to 0.30\%, below VRebalance in every scenario, and used less CPU in three of four combinations. The comparison covers complete controller configurations with different decision frequencies.
\end{abstract}

\section{Introduction}

The increasing volume and velocity of data generated by Internet of Things devices have exposed limitations of cloud-centric processing when applications require low response times or operate under constrained network bandwidth. Edge computing places computational resources closer to data sources, reducing unnecessary transfers and supporting time-sensitive processing outside centralized data centers~\cite{Shi2016EdgeView,shi2016promise}. Edge infrastructures, however, offer less computational capacity than conventional clouds and comprise heterogeneous servers and network links. Resource management is therefore central to sustaining application performance at the edge.

Distributed Stream Processing (DSP) applications continuously process unbounded data streams through directed graphs of interconnected operators. Performance depends on the interaction among input rate, operator selectivity, processing demand, queue accumulation, network communication, and the resources assigned to each task. Distributed execution supports high throughput and low latency~\cite{peng2019joint}, but an allocation suitable for one workload can become inefficient as arrival rate or application pressure changes. Underprovisioned operators accumulate queues and increase end-to-end latency, whereas overprovisioning reserves scarce edge resources that could support other applications.

Dynamic scaling adapts the resources assigned to an application at runtime. Horizontal scaling changes operator parallelism, while vertical scaling modifies the resources available to existing tasks. This work focuses on vertical scaling because it adjusts processing capacity without changing the application graph or initiating the more disruptive reconfiguration of operator parallelism. Operators have different computational demands and occupy servers with different capacities; consequently, uniform allocation changes may waste CPU or preserve local bottlenecks. VRebalance~\cite{kang2021slo}, for example, uses Bayesian Optimization (BO) for application-level vertical resource allocation, primarily guided by aggregate workload and performance information.

This work investigates whether operator-level observations allow a controller to allocate CPU selectively while maintaining application-level performance. Vertical scaling is formulated as a Partially Observable Markov Decision Process (POMDP) and controlled with Proximal Policy Optimization (PPO)~\cite{schulman2017proximal}. At one-second intervals, the policy selects relative CPU adjustments for every application task. Its reward prioritizes compliance with a 95th-percentile end-to-end latency Service Level Objective (SLO) and, while compliant, encourages lower CPU allocation and stable reconfiguration behavior. The research question is:

\begin{quote}
\emph{Can an operator-aware PPO policy balance CPU allocation efficiency with throughput preservation and p95 end-to-end latency SLO compliance in edge DSP applications?}
\end{quote}

The question is examined through three criteria: a lower mean run-level SLO violation rate than the selected baseline (H1); lower mean total CPU allocation without a higher mean SLO violation rate (H2); and profile-level throughput that follows the offered workload without a sustained deficit (H3). The contributions are a POMDP formulation of vertical scaling on heterogeneous edge infrastructures; an observation representation combining task pressure, server capacity, and application performance; a joint incremental action space for coordinated task-level adjustments; an SLO-aware reward that gives priority to compliance and recovery; and a comparison with VRebalance across two application profiles and two workloads.

The remainder of this paper is organized as follows. Section~2 reviews related work on dynamic scaling for edge stream processing. Section~3 defines the system model and vertical-scaling problem. Section~4 presents the operator-aware PPO autoscaler, including its observation space, action space, and reward. Section~5 describes the simulation environment, scenarios, baseline, and evaluation protocol. Section~6 reports and discusses the experimental results, and Section~7 concludes the paper and outlines future work.

\section{Related Work}

Existing proposals for dynamic scaling of edge stream processing use heuristics, explicit optimization, machine learning, or communication and control mechanisms. Belkhiria and Tedeschi~\cite{belkhiria2019design} proposed decentralized horizontal scaling in which operators decide from their local workload and neighboring operators. Nardelli et al.~\cite{nardelli2018multilevel} developed a hierarchical framework that coordinates operator, application, infrastructure, and region managers. These approaches adapt parallelism or the active infrastructure rather than the CPU allocation of fixed tasks.

Machine-learning approaches have also addressed elastic parallelism. Xu and Palanisamy~\cite{xu2021modelreinf} modeled operator parallelism through Markov decision processes and contextual bandits. Arkian et al.~\cite{arkian2021model} combined model-based control with geo-distributed autoscaling. Such approaches primarily target horizontal scaling and commonly optimize throughput or latency.

VRebalance~\cite{kang2021slo} is the closest baseline for this work because it modifies the CPU resources assigned to existing workers without changing their parallelism. It uses BO to explore task-allocation vectors and application-level feedback to balance p95 latency with resource cost. Its evaluation reported an 83.7\% reduction in SLO violations compared with the default Apache Storm scaling algorithm. In contrast, the proposed policy observes individual task and host conditions and selects all relative task adjustments jointly.

Across the reviewed studies, latency or throughput appeared in every evaluation, but resource consumption was not always considered. Most proposals change parallelism or placement, while the closest vertical scaler reasons from application-level feedback. The gap addressed here is whether joint task-level CPU adjustments can reduce p95 latency SLO violations while preserving throughput and avoiding unnecessary allocation on a fixed edge placement.

\section{System Model and Problem Formulation}

The infrastructure comprises heterogeneous edge servers $\mathcal{E}$ interconnected by network links. Data sources generate streams processed by a DSP application represented as a directed acyclic graph. Operators are materialized as tasks in $\mathcal{K}$. A server $\mathcal{E}_i$ has CPU capacity $c_i$, processing capacity $\widehat{c}_i$, and RAM capacity $m_i$. Task $k$ has MIPS demand $h_k$ per event, memory demand $d_k$, and time-varying reserved CPU $c_k(t)$. Its placement $x_{i,k}$ remains fixed during an execution. When task $k$ is placed on server $i$, its effective processing capacity is

\begin{equation}
\sigma_{i,k}(t)=\widehat{c}_i\frac{c_k(t)}{c_i}.
\label{eq:processing-capacity}
\end{equation}

Increasing $c_k(t)$ increases the processing capacity assigned to the task and reduces its nominal per-event processing time. Its effect on application latency may be delayed because events can already be queued locally or at downstream operators.

End-to-end latency spans event generation at the source to completion at the sink, including queueing, network, and processing time. Let $L_{95}(t)$ denote the 95th percentile of the latency of events completed during interval $t$. The SLO is satisfied when $L_{95}(t)\leq g_j$, where $g_j$ is the threshold for application profile $j$. Task input rate $\lambda_k(t)$, throughput $\phi_k(t)$, and inbound queue $q_k(t)$ represent offered work, processing progress, and accumulated pressure.

Each task receives between $c^{\min}=500$ and $c^{\max}=10000$ millicores. Allocations must obey

\begin{equation}
c^{\min}\leq c_k(t)\leq c^{\max},\qquad
\sum_k x_{i,k}c_k(t)\leq c_i,\quad \forall i.
\label{eq:cpu-constraints}
\end{equation}

Memory demand and placement are fixed; CPU is the only resource modified by the controller. This choice follows VRebalance, which excludes memory from online resource configuration because low-latency stream processing should avoid costly storage operations on the critical processing path, while exploring memory configurations online may cause thrashing or other unpredictable behavior~\cite{kang2021slo}. The control objective is hierarchical: first avoid or recover from p95 violations; then, while compliant, reduce unnecessary CPU without producing unstable reconfigurations or a sustained throughput deficit.

\section{Operator-Aware Vertical Autoscaler}

\begin{figure}[ht]
\centering
\begin{tikzpicture}[
  block/.style={draw,rounded corners=2pt,align=center,minimum height=9mm,text width=25mm,fill=blue!7,font=\footnotesize},
  action/.style={draw,rounded corners=2pt,align=center,minimum height=9mm,text width=27mm,fill=green!9,font=\footnotesize},
  flow/.style={-{Latex[length=2mm]},thick},
  node distance=8mm and 10mm]
\node[block] (sim) {EdgeStreamPy\\DSP execution};
\node[block,right=of sim] (obs) {Normalized\\observation};
\node[block,right=of obs] (policy) {PPO policy\\joint decision};
\node[action,below=of policy] (target) {Target CPU deltas\\per task};
\node[action,below=of obs] (alloc) {Constraint-aware\\CPU allocation};
\draw[flow] (sim) -- node[above,font=\scriptsize] {metrics} (obs);
\draw[flow] (obs) -- (policy);
\draw[flow] (policy) -- (target);
\draw[flow] (target) -- (alloc);
\draw[flow] (alloc) -| node[pos=.25,below,font=\scriptsize] {allocation} (sim);
\draw[flow,dashed] (sim.north) to[bend left=18] node[above,font=\scriptsize] {training reward} (policy.north);
\end{tikzpicture}
\caption{One-second observation, decision, and vertical-scaling control loop.}
\label{fig:control-loop}
\end{figure}
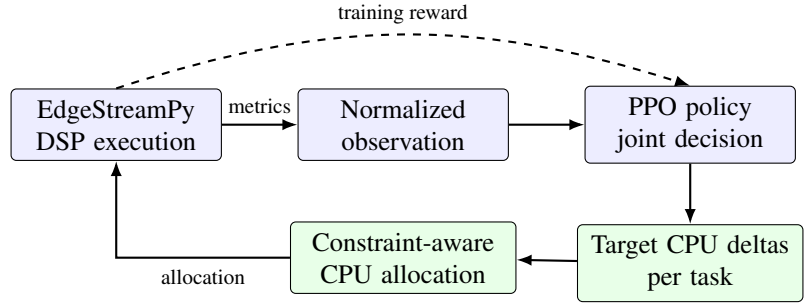

The autoscaler adjusts the CPU reserved for every task while preserving application topology and placement. At each interval, the agent receives a structured observation with task-, server-, and application-level information and selects a joint action containing one relative CPU adjustment per task. After the action is constrained by task and server capacities, EdgeStreamPy advances the application for one simulated second and returns the next observation and reward. During training, transitions from multiple simulator instances update the PPO policy. During evaluation, the policy operates deterministically and its parameters remain fixed. Figure~\ref{fig:control-loop} summarizes this interaction.

\subsection{POMDP and Observation Space}

The interaction is represented by the POMDP $\langle\mathcal{S},\mathcal{A},T,R,\Omega,O,\gamma\rangle$. The latent state contains the internal configuration required by the simulator dynamics, including workload progress, event queues and timestamps, network transfers, resource allocations, metrics, and temporal-history accumulators. PPO does not receive that state directly. A deterministic observation function constructs $o_t=O(s_t)$ from selected metrics, the policy samples or selects $\mathbf{a}_t$ from $\pi_\theta(\mathbf{a}_t\mid o_t)$, and a reward is produced after one second of simulated execution.

The observation is a dictionary of three flattened vectors,

\begin{equation}
o_t=\{\mathbf{o}^{\mathrm{task}}_t,\mathbf{o}^{\mathrm{server}}_t,\mathbf{o}^{\mathrm{app}}_t\}.
\label{eq:observation}
\end{equation}

There are 15 features per task, five per server, and 11 for the application. Table~\ref{tab:observation-features} groups these features by control purpose and normalization.

\begin{table}[ht]
\centering
\footnotesize
\caption{Observation features, normalization, and control purpose.}
\label{tab:observation-features}
\begin{tabular}{|p{.13\textwidth}|p{.25\textwidth}|p{.21\textwidth}|p{.27\textwidth}|}
\hline
\textbf{Level} & \textbf{Feature group} & \textbf{Normalization} & \textbf{Control purpose} \\ \hline
Task & CPU allocation; input rate; MIPS demand; throughput; inbound queue & CPU: clipped linear; rates, demand, and queue: $\log(1+x)$ & Represent capacity, offered work, processing demand, and accumulated pressure. \\ \hline
Task & Throughput/input ratio; queue pressure; demand/CPU & Ratios clipped to configured ranges; demand/CPU uses $\log(1+x)$ & Identify tasks whose assigned capacity is insufficient for their input. \\ \hline
Task & Task latency; processing latency; latency change; queue change & Latencies relative to $g_j$; changes clipped to $[-2,2]$ & Expose local delay and whether pressure is increasing or decreasing. \\ \hline
Task & I/O ratio; host CPU utilization; host free CPU & I/O ratio and free CPU in $[0,1]$; utilization clipped & Distinguish processing characteristics and capacity available for scale-up. \\ \hline
Server & CPU utilization/free CPU; memory utilization/free memory; task count & Capacity ratios clipped; task count normalized & Represent contention and feasibility of additional allocation on each host. \\ \hline
Application & Mean and p95 latency; SLO margin; p95 trend; violating-interval history & Latencies relative to $g_j$; margin and trend clipped & Represent global compliance, proximity to violation, and recent behavior. \\ \hline
Application & Total, mean, maximum, and minimum task CPU; total queue; throughput & CPU linearly normalized; queue and throughput use $\log(1+x)$ & Summarize resource distribution and application-level work completion. \\ \hline
\end{tabular}
\end{table}

The representation combines global performance with localized evidence of pressure. The policy can distinguish an application-wide latency increase caused by one saturated operator from a condition in which all tasks have sufficient capacity. Because tasks and servers occupy fixed positions in the flattened vectors, observation and action dimensions depend on the application graph. An independent policy is therefore trained for each profile.

\subsection{Joint Action and Feasibility}

The MultiDiscrete action contains one component per task:

\begin{equation}
\mathbf{a}_t=(a_{t,1},\ldots,a_{t,|\mathcal{K}|}),\qquad
a_{t,k}\in\{-500,-50,0,50,500\}.
\label{eq:action}
\end{equation}

Each value is a relative CPU adjustment in millicores, applied to the current allocation and clipped to the task bounds. Coarse changes respond to large pressure variations, fine changes adjust allocation near an efficient point, and zero explicitly preserves a task allocation. Components are selected jointly, allowing the policy to increase a bottleneck task while reducing CPU elsewhere.

The target vector is projected onto the feasible set defined by Equation~\ref{eq:cpu-constraints}. Reductions are processed before increases so that capacity released on a host can be reused in the same decision. If a requested increase exceeds host free CPU, the applied vector differs from the target. The next observation and reward use the allocation actually applied, preventing credit for infeasible actions.

\subsection{SLO-Aware Reward}

The reward combines seven components and is clipped to $[-6,2.5]$:

\begin{equation}
r_t=\operatorname{clip}\left(R_{\mathrm{SLO}}+R_{\mathrm{risk}}+R_{\mathrm{resource}}+
R_{\mathrm{progress}}+R_{\mathrm{churn}}+R_{\mathrm{recovery}}+R_{\mathrm{reconf}},-6,2.5\right).
\label{eq:reward}
\end{equation}

When $L_{95}(t)\leq g_j$, the agent receives a base compliance reward. A quadratic risk penalty grows between a safe-latency threshold and the SLO. Resource efficiency is rewarded as total applied CPU approaches the minimum feasible allocation, progress rewards a reduction between consecutive intervals, and churn penalizes unnecessary changes. Transitioning from a violating interval back to compliance adds a recovery bonus.

When $L_{95}(t)>g_j$, the risk, efficiency, and CPU-reduction terms are zero. The SLO term becomes a negative function of relative violation severity, recovery rewards reductions in p95 latency, and reconfiguration rewards CPU increases aligned with task pressure. Pressure is the maximum of normalized throughput deficit, queue pressure, and processing latency relative to the SLO. Increasing a task without observed pressure is discouraged, and the churn penalty is reduced to permit faster recovery. The same reward structure is used for both application profiles and workloads. Its empirically selected coefficients encode a directional SLO-first hierarchy rather than a lexicographic guarantee; isolating the components would require an ablation study.

\begin{table}[ht]
\centering
\footnotesize
\caption{Reward components and their control roles.}
\label{tab:reward-components}
\begin{tabular}{|p{.18\textwidth}|p{.15\textwidth}|p{.53\textwidth}|}
\hline
\textbf{Component} & \textbf{Range} & \textbf{Role} \\ \hline
$R_{\mathrm{SLO}}$ & $[-5,0.5]$ & Rewards compliance and penalizes violations according to relative severity. \\ \hline
$R_{\mathrm{risk}}$ & $[-0.5,0]$ & Discourages operation close to the SLO while compliant. \\ \hline
$R_{\mathrm{resource}}$ & $[0,1.5]$ & Rewards low total CPU only while latency is compliant. \\ \hline
$R_{\mathrm{progress}}$ & $[-0.25,0.25]$ & Rewards CPU reduction and penalizes movement in the opposite direction during compliance. \\ \hline
$R_{\mathrm{churn}}$ & $[-0.05,0]$ & Penalizes unnecessary changes in task allocation. \\ \hline
$R_{\mathrm{recovery}}$ & $[-1,1]$ & Rewards lower latency during violation and transition back to compliance. \\ \hline
$R_{\mathrm{reconf}}$ & $[-0.5,0.5]$ & Rewards scale-up aligned with task pressure during SLO recovery. \\ \hline
\end{tabular}
\end{table}

\section{Experimental Methodology}

\subsection{Simulation Environment and Scenarios}

The evaluation uses EdgeStreamPy, a discrete-time simulator built on EdgeSimPy~\cite{souza2023edgesimpy}. The infrastructure is an undirected network graph in which switches route flows and edge servers provide CPU capacity and processing speed. A DSP application is a DAG of simulated tasks. Each task consumes CPU, has a MIPS demand per event and a selectivity, and exchanges events through network flows when communicating with tasks on other servers. The simulator models event generation, queues, processing, network bandwidth and propagation, and CPU reallocation. Although it also supports other resource-management operations, migration and horizontal scaling are disabled here.

Each scenario contains seven heterogeneous edge servers in a compact $3\times3$ hexagonal-grid region. Adjacent servers have bidirectional 1~Gbps links with 10~ms propagation latency. Tasks are randomly placed with at most two tasks per server, and their placement remains fixed during execution. The evaluation uses PRED and ETL profiles adapted from RIoTBench~\cite{shukla2017riotbench}. PRED has six simulated tasks, including two prediction branches, whereas ETL has nine tasks in a longer chain. The simulator does not execute their business logic; profile behavior is determined by processing demand, selectivity, event routing, and allocation.

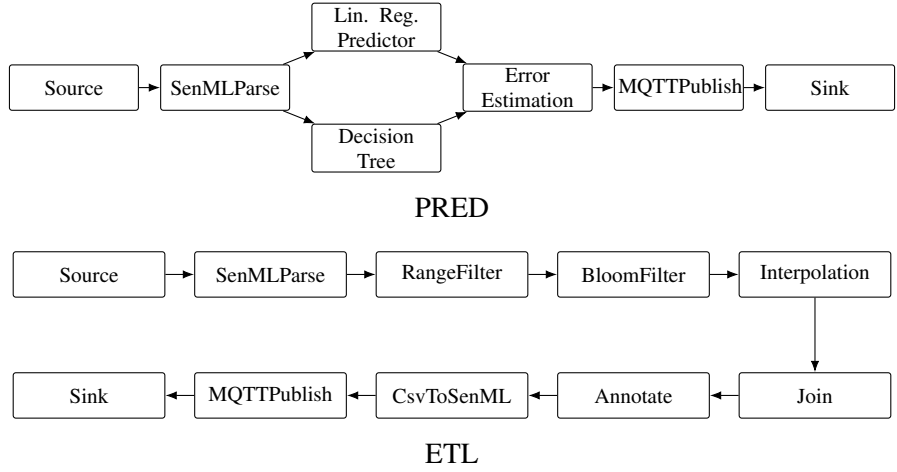
\begin{figure}[ht]
\centering

% ---------- PRED ----------
\begin{tikzpicture}[
  op/.style={draw,rounded corners=1pt,align=center,inner sep=1.5pt,text width=16mm,minimum height=6mm,font=\scriptsize},
  flow/.style={-{Latex[length=1.5mm]},thin}]
\node[op] (p1) at (0,0)      {Source};
\node[op] (p2) at (2.0,0)    {SenMLParse};
\node[op] (p3) at (4.0,0.8)  {Lin. Reg.\\Predictor};
\node[op] (p4) at (4.0,-0.8) {Decision\\Tree};
\node[op] (p5) at (6.0,0)    {Error\\Estimation};
\node[op] (p6) at (8.0,0)    {MQTTPublish};
\node[op] (ps) at (10.0,0)   {Sink};
\draw[flow] (p1) -- (p2);
\draw[flow] (p2) -- (p3);
\draw[flow] (p2) -- (p4);
\draw[flow] (p3) -- (p5);
\draw[flow] (p4) -- (p5);
\draw[flow] (p5) -- (p6);
\draw[flow] (p6) -- (ps);
\end{tikzpicture}

\smallskip
\small PRED

\bigskip

% ---------- ETL (pipeline linear, quebrado em 2 linhas) ----------
\begin{tikzpicture}[
  op/.style={draw,rounded corners=1pt,align=center,inner sep=1.5pt,text width=19mm,minimum height=6mm,font=\scriptsize},
  flow/.style={-{Latex[length=1.5mm]},thin}]
% Linha 1: esquerda -> direita
\node[op] (e1) at (0,0)   {Source};
\node[op] (e2) at (2.4,0) {SenMLParse};
\node[op] (e3) at (4.8,0) {RangeFilter};
\node[op] (e4) at (7.2,0) {BloomFilter};
\node[op] (e5) at (9.6,0) {Interpolation};

% Linha 2: direita -> esquerda
\node[op] (e6) at (9.6,-1.6) {Join};
\node[op] (e7) at (7.2,-1.6) {Annotate};
\node[op] (e8) at (4.8,-1.6) {CsvToSenML};
\node[op] (e9) at (2.4,-1.6) {MQTTPublish};
\node[op] (es) at (0,-1.6)   {Sink};

\draw[flow] (e1) -- (e2);
\draw[flow] (e2) -- (e3);
\draw[flow] (e3) -- (e4);
\draw[flow] (e4) -- (e5);
\draw[flow] (e5) -- (e6);
\draw[flow] (e6) -- (e7);
\draw[flow] (e7) -- (e8);
\draw[flow] (e8) -- (e9);
\draw[flow] (e9) -- (es);
\end{tikzpicture}

\smallskip
\small ETL

\caption{PRED and ETL application topologies adapted from RIoTBench.}
\label{fig:application-topologies}
\end{figure}

Two arrival-rate traces are replayed. TAXI comes from the smart-transportation stream used by RIoTBench and contributes the first 3,600 points of a trace beginning on January 1, 2010. INTEL comes from measurements collected by 54 sensors in the Intel Berkeley Research laboratory~\cite{intel_lab_data_2004}. It contains 1,056 points and is replayed cyclically until the 3,600-interval horizon is complete. In both traces, source data are aggregated into five-minute windows and each point is replayed during one simulator second. Thus, a run comprises 3,600 one-second intervals and represents 60 simulated minutes. The INTEL cycles are repeated observations of the constructed periodic workload, not independent experimental replications.

The generic SLO is instantiated separately for each profile: 240~ms for ETL and 180~ms for PRED. ETL uses a higher latency reference because its operator chain is longer than PRED's. The same profile-specific thresholds are applied to both controllers under both workloads. The SLO violation rate is the percentage of one-second intervals in which p95 end-to-end latency exceeds the profile threshold. Throughput is the number of events reaching the sink during the interval, and allocation is the sum of CPU reserved for all profile tasks.

\subsection{Baseline, Training, and Protocol}

VRebalance was reimplemented for EdgeStreamPy. Its BO procedure uses application p95 latency and allocated CPU to score an allocation vector, while sink throughput selects a workload tier. A separate optimizer and observation history are kept per tier, and the same exploration window of 5 to 16 minutes is used. All tasks start at 500 millicores. The first allocation is held for 60 simulator steps, after which the optimizer is evaluated once every 60 steps. Candidate allocations use the same 500--10,000 millicore task bounds as PPO and are quantized in 50-millicore increments (the only parameter altered from original solution is the allocation bounds, to reflect the new infrastructure). When requests exceed a server's capacity, each task retains the minimum and remaining CPU is distributed according to requested demand above that minimum.

Independent PPO models were trained for PRED and ETL because their observation and action dimensions differ. The training workload was a 352-point segment derived from the penultimate day of the taxi dataset and distinct from the final TAXI trace. The segment was selected to reduce the wall-clock cost of training and intermediate evaluations while retaining gradual, non-monotonic increases and decreases that exercise both scale-up and scale-down decisions. The original trace and a copy with doubled arrival rates were combined with placements generated from seeds 42 and 37, producing four training scenarios per profile. These variations followed preliminary single-placement runs that did not generalize to different task-to-server mappings; together, they expose the policy to changes in base communication latency, server capacity, and processing pressure. Each model received a budget of 500,000 aggregate environment steps with 14 parallel simulator instances and stopped after 516,096 transitions. Both used Stable-Baselines3 PPO with a \texttt{MultiInputPolicy} and separate actor and critic networks with two hidden layers of 64 units. Table~\ref{tab:ppo-configuration} reports the remaining configuration. One fixed policy was retained per profile based on mean reward over development episodes.

\begin{table}[ht]
\centering
\caption{PPO training configuration used for both application profiles.}
\label{tab:ppo-configuration}
\begin{tabular}{lc@{\hspace{1cm}}lc}
\hline
Learning rate & $3\times10^{-4}$ & Steps/rollout/environment & 2,048 \\
Minibatch size & 256 & Optimization epochs & 20 \\
Discount factor $\gamma$ & 0.995 & GAE factor $\lambda$ & 0.95 \\
Policy clipping & 0.20 & Value-function clipping & 0.20 \\
Entropy coefficient & 0.05 & Value-function coefficient & 0.50 \\
Maximum gradient norm & 0.50 & Training seed & 284572 \\
\hline
\end{tabular}
\end{table}

The two profiles and two workloads yield four combinations. Ten distinct random task placements are evaluated for each combination and controller, producing 40 runs per controller and 80 total. Every PPO run is paired with a VRebalance run using the same workload, placement, and evaluation seed. Evaluation placements differ from training placements. PPO is deterministic and reuses one policy per profile; consequently, the ten repetitions measure placement variability rather than independent training variability.

The comparison covers complete controller configurations. PPO selects an allocation after every one-second monitoring interval, whereas VRebalance updates once every 60 seconds. For H3, each run's throughput is its arithmetic mean over 3,600 intervals. Each of the ten matched PPO--VRebalance pairs yields one relative difference, $(\bar{\phi}_{\mathrm{PPO}}-\bar{\phi}_{\mathrm{VRebalance}})/\bar{\phi}_{\mathrm{VRebalance}}\times100$. A 95\% percentile-bootstrap confidence interval for the mean paired difference is computed from 10,000 resamples of the ten complete pairs. PPO is considered non-inferior when the lower confidence bound exceeds the $-5\%$ margin. The repeated-run analysis reports latency SLO mitigation (H1), CPU allocation under compliance (H2), and throughput preservation (H3).

\section{Experimental Results}

\subsection{Tail Latency and SLO Mitigation}

Across the repeated paired executions, PPO shows more consistent p95 latency compliance than VRebalance in all four combinations. The following analyses distinguish temporal behavior, run-level means, and SLO violation rates.

\begin{figure}[!htbp]
\centering
\includegraphics[width=.9\textwidth]{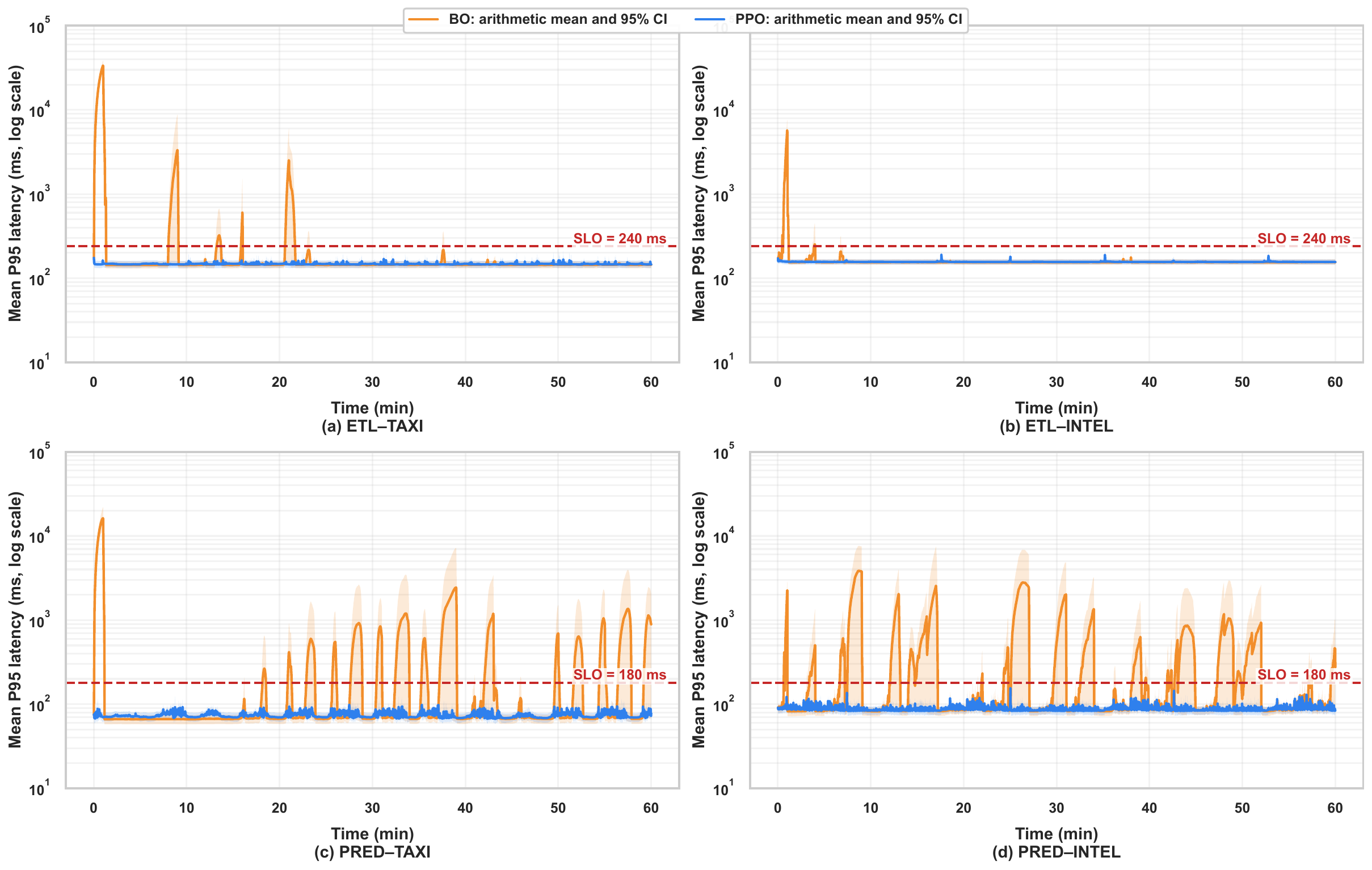}
\caption{P95 end-to-end latency over time. Lines are pointwise arithmetic means for the ten executions.}
\label{fig:latency-timeseries}
\end{figure}

Figure~\ref{fig:latency-timeseries} shows the temporal behavior behind these run-level summaries. The pointwise mean p95 latency of PPO remains below the corresponding SLO throughout execution in all four combinations. VRebalance presents repeated excursions above the thresholds, with more pronounced variation among placements for the PRED profile.

\begin{figure}[!htbp]
\centering
\includegraphics[width=.9\textwidth]{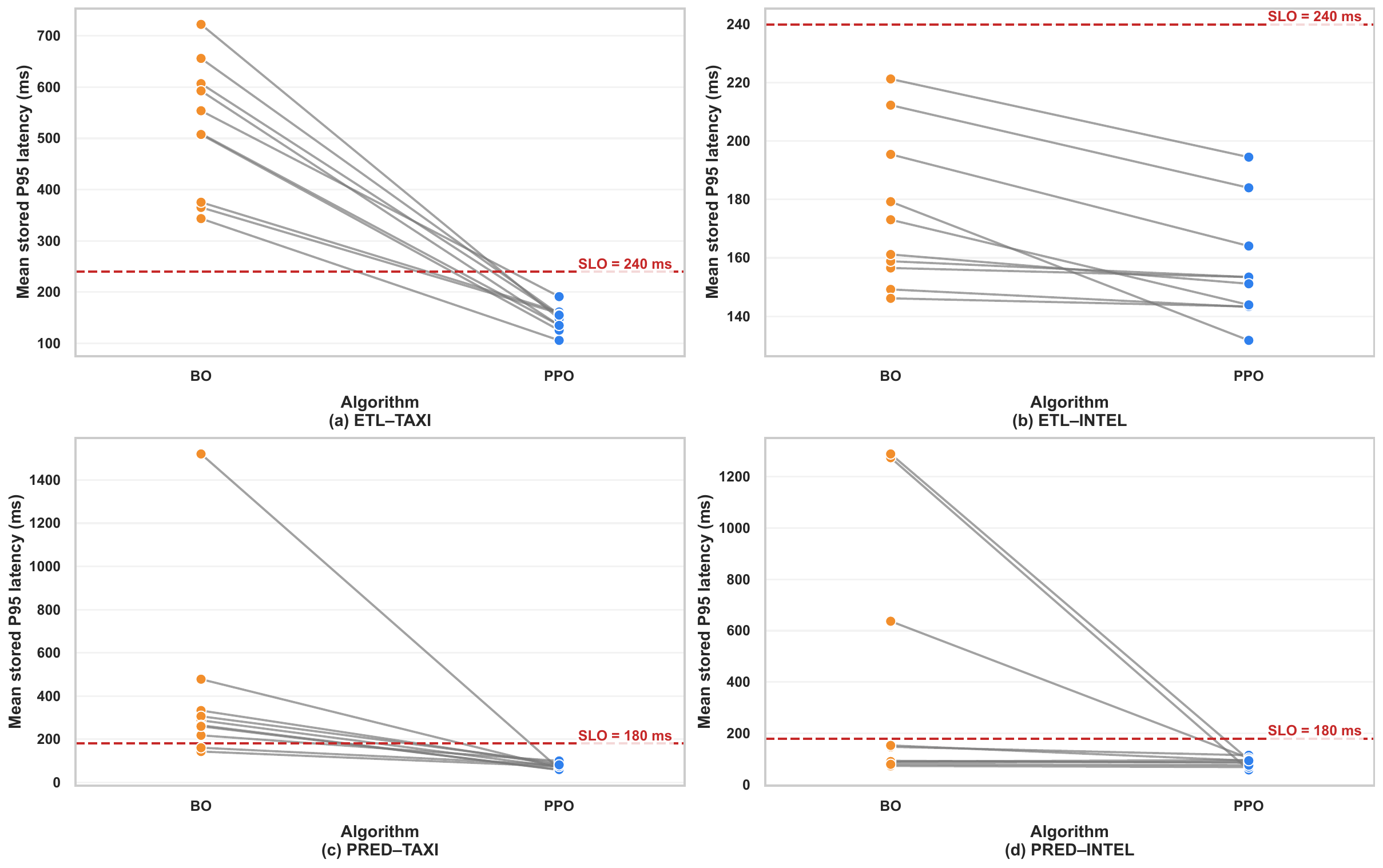}
\caption{Paired per-execution mean p95 end-to-end latency. Dashed horizontal lines are the profile-specific SLOs.}
\label{fig:paired-latency}
\end{figure}

The paired run-level means in Figure~\ref{fig:paired-latency} show that this group pattern is not solely a consequence of averaging compliant and non-compliant executions. Every PPO execution has a mean p95 latency below its profile-specific SLO. For VRebalance, ETL--INTEL is the only scenario in which every run-level mean remains below the threshold. ETL--TAXI presents higher run means despite a lower violation rate than the PRED scenarios because a small number of extreme intervals strongly affect the mean.

\begin{figure}[!htbp]
\centering
\includegraphics[width=.87\textwidth]{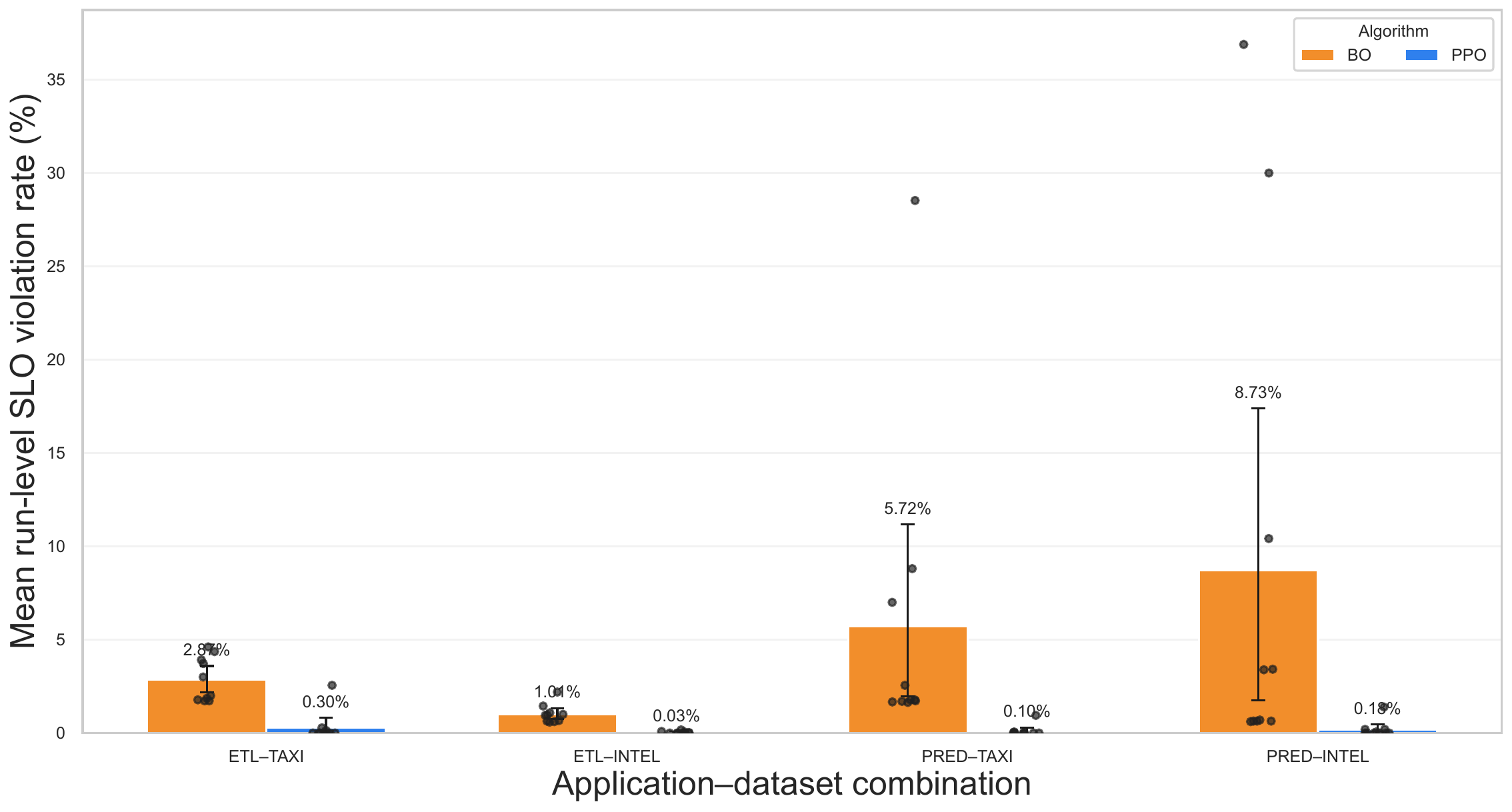}
\caption{Run-level p95 latency SLO violation rates. Bars are means of ten run-level rates, points are executions, and whiskers are 95\% percentile confidence interval.}
\label{fig:slo-violations}
\end{figure}

Figure~\ref{fig:slo-violations} reports the run-level violation rates. PPO obtains mean rates of 0.30\%, 0.03\%, 0.10\%, and 0.18\% for ETL--TAXI, ETL--INTEL, PRED--TAXI, and PRED--INTEL. The corresponding VRebalance means are 2.87\%, 1.01\%, 5.72\%, and 8.73\%. The descriptive difference ranges from 0.98 percentage points in ETL--INTEL to 8.55 in PRED--INTEL. The latter also includes baseline runs above 30\%, exposing placement sensitivity hidden by the mean. Under the predefined criterion, H1 is met in every combination.

\subsection{CPU Allocation Efficiency Under Compliance}

\begin{figure}[!htbp]
\centering
\includegraphics[width=.87\textwidth]{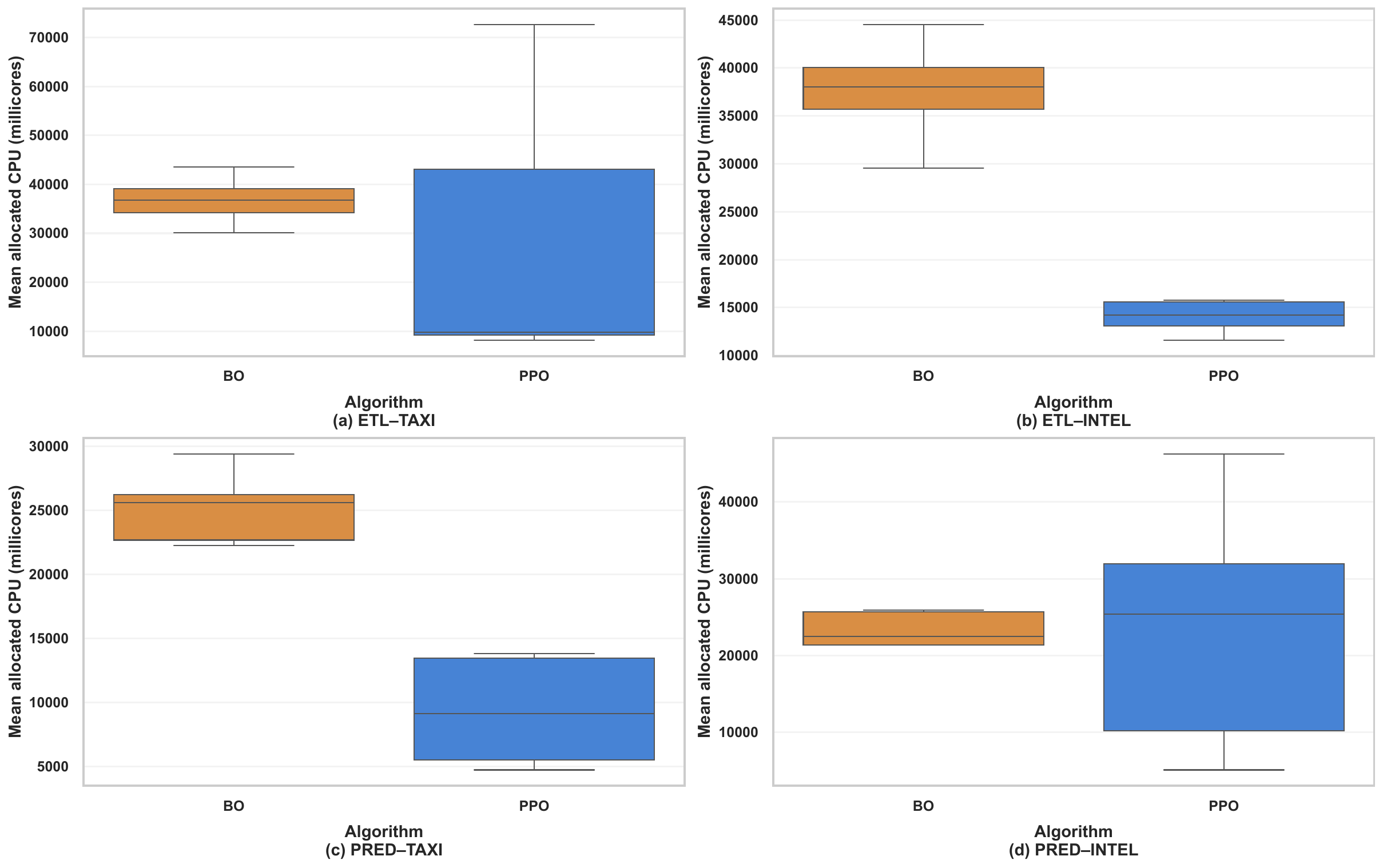}
\caption{Distribution of per-execution mean total CPU allocation. The center line is the median.}
\label{fig:cpu-allocation}
\end{figure}

H2 concerns CPU efficiency while preserving the SLO behavior established under H1, rather than minimum allocation in isolation. Figure~\ref{fig:cpu-allocation} shows lower CPU under PPO in three scenarios. For ETL--TAXI, ETL--INTEL, and PRED--TAXI, mean and median allocations are lower, with central-tendency differences of approximately 15,000--25,000 millicores. PPO allocation remains below VRebalance for more than 80\% of the horizon in these scenarios and follows workload changes at its one-second control interval.

PRED--INTEL is the exception: PPO's median allocation is slightly higher. In that scenario, however, PPO has a mean violation rate of 0.18\%, compared with 8.73\% for VRebalance, the baseline's highest rate among the combinations. This joint result is consistent with the reward hierarchy: the policy accepts more CPU when the simulated profile is under greater SLO pressure. It does not demonstrate that additional CPU alone caused the lower violation rate. H2 is therefore not supported uniformly; the evidence indicates a scenario-dependent tendency toward resource savings.

\subsection{Throughput Preservation}

\begin{figure}[!htbp]
\centering
\includegraphics[width=.75\textwidth]{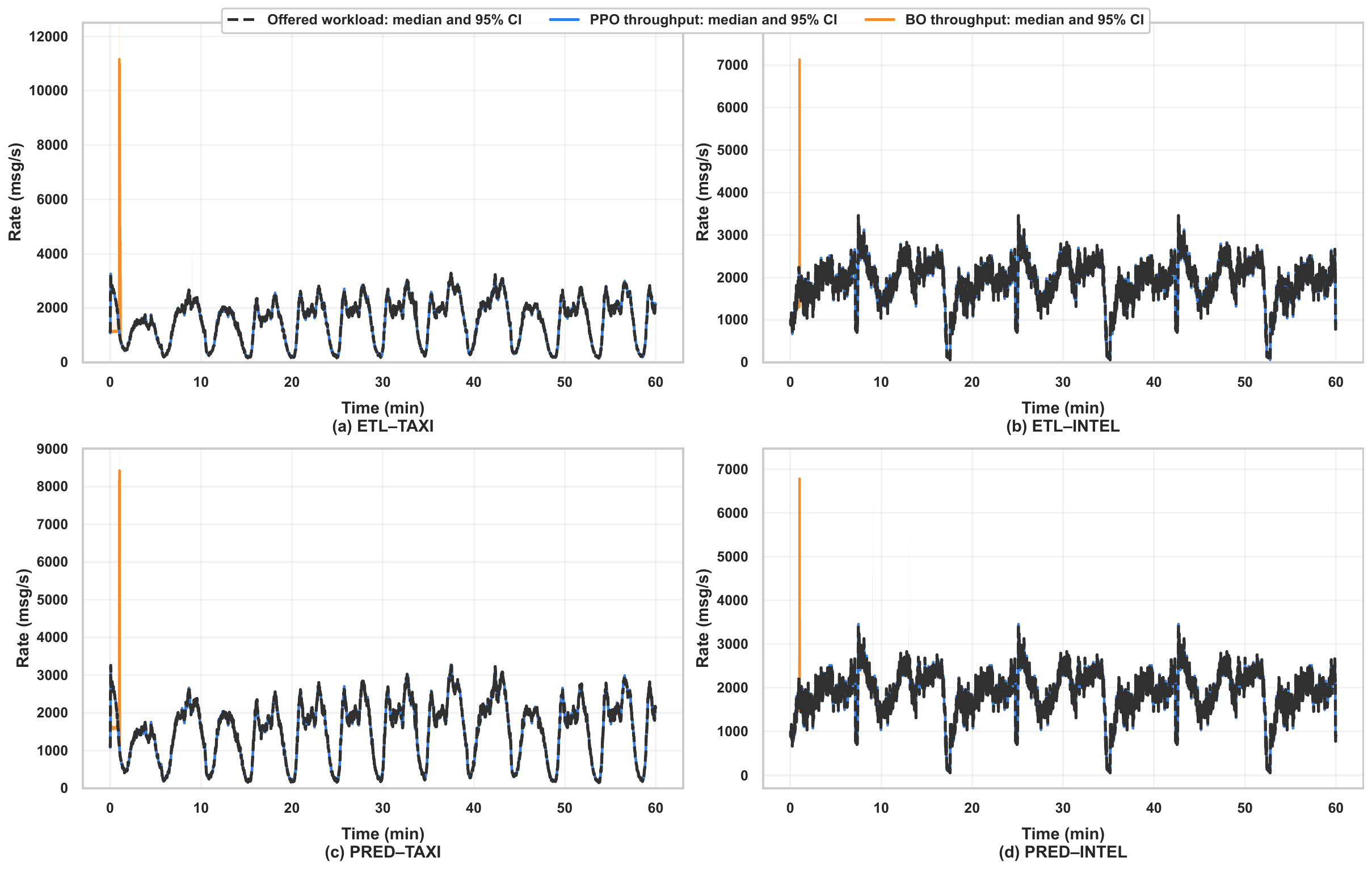}
\caption{Offered workload and achieved throughput for the four profile--workload combinations.}
\label{fig:throughput}
\end{figure}

Figure~\ref{fig:throughput} compares offered workload and achieved throughput. After the initial transient, the median PPO and VRebalance curves are nearly indistinguishable from the offered-rate trajectory, and their confidence intervals remain closely aligned for almost the entire horizon. PPO shows no sustained throughput deficit.

With a non-inferiority margin of $-5\%$, PPO is non-inferior to VRebalance in all four combinations. The mean paired differences (PPO minus baseline) are approximately 0.001\% for ETL--TAXI and ETL--INTEL, 0.026\% for PRED--TAXI (95\% CI: $-0.005\%$ to 0.069\%), and $-0.001\%$ for PRED--INTEL (95\% CI: $-0.003\%$ to 0.001\%). All lower confidence bounds are above the margin, and all differences remain within a $\pm5\%$ practical-equivalence band. H3 is therefore supported in every scenario.

\section{Conclusion and Future Work}

This work investigated whether an operator-aware DRL controller can reduce CPU allocation for edge DSP applications while preserving throughput and p95 end-to-end latency SLO compliance relative to VRebalance. Vertical autoscaling was formulated as a POMDP and controlled with PPO, which jointly adjusts per-task CPU from task, server, and application-level observations and prioritizes SLO compliance and recovery before resource reduction.

Across the evaluated profiles, workloads, and paired placements, the results provide a qualified affirmative answer to the research question. PPO reduced mean SLO violation rates to 0.03--0.30\%, compared with 1.01--8.73\% for VRebalance, supporting H1 in all four combinations. It also preserved throughput and satisfied the non-inferiority criterion throughout the evaluation, supporting H3. CPU savings were scenario-dependent: PPO used less CPU in three combinations, whereas PRED--INTEL required slightly more CPU while reducing the mean violation rate from 8.73\% to 0.18\%. H2 therefore received mixed support, reflecting an SLO-first trade-off in which resource reduction depends on application and workload conditions.

Future work should test independent training seeds, additional profiles, workloads, infrastructure topologies, and placements. Component ablations should vary observation groups, reward terms, action increments, and decision frequency to separate their effects. Finally, the controller should be evaluated on a stream-processing engine in emulated and physical edge environments, including measurements of inference and actuation overheads, runtime effects, metric noise, and reallocation cost.

\clearpage
\bibliographystyle{sbc}
\bibliography{sbc-template}

@article{Shi2016EdgeView,
  author  = {Shi, Weisong and Cao, Jie and Zhang, Quan and Li, Youhuizi and Xu, Lanyu},
  title   = {Edge Computing: Vision and Challenges},
  journal = {IEEE Internet of Things Journal},
  year    = {2016},
  volume  = {3},
  number  = {5},
  pages   = {637--646},
  doi     = {10.1109/JIOT.2016.2579198}
}

@article{shi2016promise,
  title     = {The Promise of Edge Computing},
  author    = {Shi, Weisong and Dustdar, Schahram},
  journal   = {Computer},
  volume    = {49},
  number    = {5},
  pages     = {78--81},
  year      = {2016},
  publisher = {IEEE}
}

@inproceedings{peng2019joint,
  title        = {Joint Operator Scaling and Placement for Distributed Stream Processing Applications in Edge Computing},
  author       = {Peng, Qinglan and Xia, Yunni and Wang, Yan and Wu, Chunrong and Luo, Xin and Lee, Jia},
  booktitle    = {Service-Oriented Computing: 17th International Conference, ICSOC 2019},
  pages        = {461--476},
  year         = {2019},
  organization = {Springer}
}

@inproceedings{kang2021slo,
  title        = {SLO-Aware Virtual Rebalancing for Edge Stream Processing},
  author       = {Kang, Peng and Lama, Palden and Khan, Samee U.},
  booktitle    = {2021 IEEE International Conference on Cloud Engineering (IC2E)},
  pages        = {126--135},
  year         = {2021},
  organization = {IEEE}
}

@inproceedings{belkhiria2019design,
  title        = {Design and Evaluation of Decentralized Scaling Mechanisms for Stream Processing},
  author       = {Belkhiria, Mehdi Mokhtar and Tedeschi, C{\'e}dric},
  booktitle    = {2019 IEEE International Conference on Cloud Computing Technology and Science (CloudCom)},
  pages        = {247--254},
  year         = {2019},
  organization = {IEEE}
}

@inproceedings{nardelli2018multilevel,
  title        = {A Multi-Level Elasticity Framework for Distributed Data Stream Processing},
  author       = {Nardelli, Matteo and Russo Russo, Gabriele and Cardellini, Valeria and Lo Presti, Francesco},
  booktitle    = {European Conference on Parallel Processing},
  pages        = {53--64},
  year         = {2018},
  organization = {Springer}
}

@inproceedings{xu2021modelreinf,
  title        = {Model-Based Reinforcement Learning for Elastic Stream Processing in Edge Computing},
  author       = {Xu, Jinlai and Palanisamy, Balaji},
  booktitle    = {2021 IEEE 28th International Conference on High Performance Computing, Data, and Analytics (HiPC)},
  pages        = {292--301},
  year         = {2021},
  organization = {IEEE}
}

@inproceedings{arkian2021model,
  title        = {Model-Based Stream Processing Auto-Scaling in Geo-Distributed Environments},
  author       = {Arkian, HamidReza and Pierre, Guillaume and Tordsson, Johan and Elmroth, Erik},
  booktitle    = {2021 International Conference on Computer Communications and Networks (ICCCN)},
  pages        = {1--10},
  year         = {2021},
  organization = {IEEE}
}

@article{souza2023edgesimpy,
  title     = {EdgeSimPy: Python-Based Modeling and Simulation of Edge Computing Resource Management Policies},
  author    = {Souza, Paulo S. and Ferreto, Tiago and Calheiros, Rodrigo N.},
  journal   = {Future Generation Computer Systems},
  volume    = {148},
  pages     = {446--459},
  year      = {2023},
  publisher = {Elsevier}
}

@article{shukla2017riotbench,
  title     = {RIoTBench: An IoT Benchmark for Distributed Stream Processing Systems},
  author    = {Shukla, Anshu and Chaturvedi, Shilpa and Simmhan, Yogesh},
  journal   = {Concurrency and Computation: Practice and Experience},
  volume    = {29},
  number    = {21},
  pages     = {e4257},
  year      = {2017},
  publisher = {Wiley Online Library}
}

@misc{intel_lab_data_2004,
  author       = {Bodik, Peter and Hong, Wei and Guestrin, Carlos and Madden, Samuel and Paskin, Mark and Thibaux, Romain},
  title        = {Intel Lab Data},
  year         = {2004},
  howpublished = {\url{https://db.csail.mit.edu/labdata/labdata.html}},
  note         = {Data collected from 54 sensors deployed in the Intel Berkeley Research lab}
}

@article{schulman2017proximal,
  title   = {Proximal Policy Optimization Algorithms},
  author  = {Schulman, John and Wolski, Filip and Dhariwal, Prafulla and Radford, Alec and Klimov, Oleg},
  journal = {arXiv preprint arXiv:1707.06347},
  year    = {2017}
}

\end{document}